# Survival of the Un-fittest: Why the Worst Infrastructure Gets Built -- and What We Can Do About It

By Bent Flyvbjerg, Saïd Business School, University of Oxford[*]

Article for special issue on "Infrastructure, Utilities, and Regulation" of *Oxford Review of Economic Policy* edited by Dieter Helm

Draft 3.0: Please do not circulate, copy, or quote

**ABSTRACT**
The article first describes characteristics of major infrastructure projects. Second, it documents a much neglected topic in economics: that ex ante estimates of costs and benefits are often very different from actual ex post costs and benefits. For large infrastructure projects the consequence is cost overruns, benefit shortfalls, and the systematic underestimation of risks. Third, implications for cost-benefit analysis are described, including that such analysis is not to be trusted for major infrastructure projects. Fourth, the article uncovers the causes of this state of affairs in terms of perverse incentives that encourage promoters to underestimate costs and overestimate benefits in the business cases for their projects. But the projects that are made to look best on paper are the projects that amass the highest cost overruns and benefit shortfalls in reality. The article depicts this situation as "survival of the un-fittest." Fifth, the article sets out to explain how the problem may be solved, with a view to arriving at more efficient and more democratic projects, and avoiding the scandals that often accompany major infrastructure investments. Finally, the article identifies current trends in major infrastructure development. It is argued that a rapid increase in stimulus spending combined with more investments in emerging economies combined with more spending on information technology is catapulting infrastructure investment from the frying pan into the fire.

**I. INTRODUCTION**
In mid-2008, *The Economist* called current spending on infrastructure the "biggest investment boom in history" (*The Economist*, June 7, 2008: 80). Spending was the largest it had ever been as a share of world GDP. When the fiscal crisis deepened and became global during Fall-Winter 2008-2009, this could have ended the infrastructure boom, as banks and capital funds radically cut back on their lending. But the opposite appears to have happened. Reductions in private funds have been offset by hundreds of billions of public dollars for stimulus spending. Heads of state, led by US president Barach Obama and China Premier Wen Jiabao, have singled out investment in infrastructure as a key means to create jobs and keep the economy from slumping. China was first mover when its State Council, in November 2008, passed a $586 billion stimulus plan, mainly for investment in infrastructure. In February 2009 the US followed suit, with Congress passing President Obama's $787 billion new New Deal. India has a $475 billion plan, and the UK, Germany, France, and many other nations have made similar arrangements.

---

With so much money in the pipeline--and with the health of the global economy riding on the success of infrastructure investment--the efficiency of infrastructure delivery is particularly important at present. If done right the investment boom could become a boon, because infrastructure investment is appealing in many ways: it creates and sustains employment; there is a large element of domestic inputs relative to imports; it improves productivity and competitiveness by lowering producer costs; it benefits consumers through higher-quality services; and it improves the environment when infrastructures that are environmentally sound substitute infrastructures that are not (Helm 2008: 1).

But there is a big "if" here. Because if done wrong the thrust may become a bust, with boondoggles worse than any seen yet, weakening the economy instead of improving it. Unfortunately the conventional way of delivering major infrastructure shows a dismal performance record. In what follows, I will document the record, I will explain why it is so poor, and finally I will describe measures that may help current stimulus spending become effective, instead of adding to the financial and economic failures that litter the field of infrastructure investment.

But first, let's see what are the characteristics of major infrastructure.

## II. CHARACTERISTICS OF MAJOR INFRASTRUCTURE

Major infrastructure projects generally have the following characteristics:[1]

- Such projects are inherently risky due to long planning horizons and complex interfaces.
- Technology and design are often non-standard.
- Decision making, planning, and management are typically multi-actor processes with conflicting interests.
- Often there is "lock in" or "capture" of a certain project concept at an early stage, leaving alternatives analysis weak or absent.
- The project scope or ambition level will typically change significantly over time.
- Statistical evidence shows that such unplanned events are often unaccounted for, leaving budget and time contingencies sorely inadequate.
- As a consequence, misinformation about costs, benefits, and risks is the norm throughout project development and decision-making, including in the business case.
- The result is cost overruns and/or benefit shortfalls during project implementation.

Cost overruns in the order of 50 percent in real terms are common for major infrastructure, and overruns above 100 percent are not uncommon. Demand and benefit forecasts that are wrong by 20 percent to 70 percent compared with actual development are common.

Table 1 shows more detailed cost data for transportation infrastructure projects. Transportation is used as an example here and elsewhere in the article because the best data exist for transportation and because there is not enough space to present data for all project types. It should be mentioned, however, that comparative research shows that the problems identified for transportation apply to a wide range of other project types including ICT systems, buildings, aerospace projects, defense, mega-events like the Olympics and the World Cup, water projects, dams, power plants, oil and gas extraction projects, mining, large-scale manufacturing, big science, and urban and regional development projects (Flyvbjerg, Bruzelius, and Rothengatter, 2003: 18-19; Altshuler and Luberoff, 2003; Priemus, Flyvbjerg, and van Wee, 2008 eds.; Flyvbjerg, Holm, and Buhl, 2002: 286; Flyvbjerg, 2005a).

The data set in Table 1 shows cost overrun in 258 projects in 20 nations on five continents. All projects for which data was obtainable were included in the study.[2] For rail, average cost overrun is 44.7 percent measured in constant prices from the build decision. For bridges and tunnels, the equivalent figure is 33.8 percent, and for roads 20.4 percent. The difference in cost overrun between the three project types is statistically significant (Flyvbjerg, Holm, and Buhl, 2002). The large standard deviations shown in Table 1 are as interesting as the large average cost overruns. The size of the standard deviations demonstrate that uncertainty and risk regarding cost overruns in infrastructure are large, indeed.

The following key observations pertain to cost overruns in transportation infrastructure projects:

- 9 out of 10 projects have cost overrun.
- Overrun is found across the 20 nations and 5 continents covered by the study.
- Overrun is constant for the 70-year period covered by the study, cost estimates have not improved over time.

[Table 1 app. here]

Table 2 shows the inaccuracy of travel demand forecasts for rail and road infrastructure. The demand study covers 208 projects in 14 nations on five continents. All projects for which data was obtainable were included in the study.[3] For rail, actual passenger traffic is 51.4 percent lower than estimated traffic on average. This is equivalent to an average overestimate in rail passenger forecasts of no less than 105.6 percent. The result is large benefit shortfalls for rail. For roads, actual vehicle traffic is on average 9.5 percent higher than forecasted traffic. We see that rail passenger forecasts are biased, whereas this is less the case for road traffic forecasts. The difference between rail and road is statistically significant at a high level. Again the standard deviations are large, indicating that forecasting errors vary widely across projects (Flyvbjerg, Holm, and Buhl, 2005; Flyvbjerg, 2005b).

The following observations hold for traffic demand forecasts:

- 84 percent of rail passenger forecasts are wrong by more than ±20 percent.
- 9 out of 10 rail projects have overestimated traffic.
- 50 percent of road traffic forecasts are wrong by more than ±20 percent.
- The number of roads with overestimated and underestimated traffic, respectively, is about the same.
- Inaccuracy in traffic forecasts are found in the 14 nations and 5 continents covered by the study.
- Inaccuracy is constant for the 30-year period covered by the study, forecasts have not improved over time.

We conclude that if techniques and skills for arriving at accurate cost and traffic forecasts have improved over time, these improvements have not resulted in an increase in the accuracy of forecasts.

[Table 2 app. here]

We also conclude that cost overruns and benefit shortfalls are a problem because: (1) they lead to a Pareto-inefficient allocation of resources, i.e., waste; (2) they lead to delays and further cost overruns and benefit shortfalls; (3) they destabilize project management; and (4) the problem is getting bigger, because projects get bigger.

**III. IMPLICATIONS FOR COST-BENEFIT ANALYSIS**
Cost-benefit analyses and social and environmental impact assessments are typically at the core of documentation and decision making for major infrastructure projects. Such analyses are based on cost and traffic forecasts like those described above. But if we combine the data in tables 1 and 2, we see that for rail an average cost overrun of 44.7 percent combines with an average traffic shortfall of 51.4 percent.[4] For roads, an average cost overrun of 20.4 percent combines with a fifty-fifty chance that traffic is also wrong by more than 20 percent. With errors and biases of such magnitude in the forecasts that form basis for cost-benefit analyses, such analyses will also, with a high degree of certainty, be strongly misleading. 'Garbage in, garbage out', as the saying goes.

As a case in point, consider the Channel tunnel, the longest underwater rail tunnel in Europe, connecting France and the UK. This project was sold as highly beneficial both financially and economically. At the initial public offering, Eurotunnel, the private owner of the tunnel, lured investors by telling them that 10 percent "would be a reasonable allowance for the possible impact of unforeseen circumstances on construction costs."[5] In fact, costs went 80 percent over budget for construction and 140 percent for financing, measured in real terms from the decision date. Revenues have been half of those forecasted. As a consequence the project has proved non-viable, with an internal rate of return on the investment that is negative, at minus 14.5 percent. However convenient for the users of the service--who are heavily subsidized--the Channel tunnel detracts from the economy instead of adding to it. An economic and financial ex-post evaluation of the project, which systematically compared actual with forecasted costs and benefits, concluded that "the British Economy would have been better off had the Tunnel never been constructed" (Anguera, 2006: 291).

But perhaps the Channel tunnel was just a piece of bad luck? Perhaps the next similar project did better? Not so. The next major tunnel was the Danish Great Belt rail tunnel, the second-longest underwater rail tunnel in Europe, opened three years after the Channel tunnel. Here construction cost overrun was 120 percent in real terms, and the project proved non-viable even before it opened. Only by cross-subsidizing the tunnel with revenues from a nearby motorway bridge was it possible to pay for the tunnel. Danish legislators--and the government and the EU Commission--are officially against cross subsidies. But in this case they were happy and quick to use the remedy to sweep a major embarrassment under the carpet: the fact that actual benefits would never come even close to covering actual costs for the tunnel, despite that taxpayers had been told the exact opposite when they were asked to underwrite the billion-dollar debt of the project.

The Channel and Great Belt tunnels are both examples of a main mechanism I will focus on below: "survival of the un-fittest." From an economic point of view the projects should never have been built, at least not in the form they were. They survived because benefit-cost ratios presented to investors and legislators were hugely inflated, deliberately or not. The UK, France, and Denmark are rich countries and can afford to build financial and economic disasters like this. But it is important to understand that countries do not become rich by doing so. They do so when they have become rich.

It is easy to make long lists of projects with similar cost and/or benefit problems (Flyvbjerg, 2005a). To mention only a few, in America Boston's Big Dig, LA's subway, San Francisco's Bay Bridge, Denver's new International Airport, and the New Woodrow Wilson Bridge in Washington, DC come to mind, not to speak of the rebuilding of Iraq. In Britain, the London Tube public-private partnership, the West Coast Main Line upgrade, the Railtrack fiscal collapse, the Millennium Dome, the Scottish parliament building, the Humber Bridge, and the cost overruns on the 2012 London Olympics have been major boondoggles. Other nations, and even cities, have their own long lists of underperforming projects.

This does not show the uselessness of cost-benefit analysis as such, needless to say. But if informed decisions are the goal, then conventional ex ante cost-benefit analysis must be supplemented with empirical ex post risk analysis focused on documented uncertainties in the estimates of costs and benefits that enter into cost-benefit analysis. For a given major infrastructure project, this would constitute a kind of empirical due diligence of its cost-benefit analysis, something that is rarely carried out today.

Given the data presented above, a key recommendation for decision makers, investors, and voters who care about what Williams (1998) calls "honest numbers" is they should not trust the budgets, patronage forecasts, and cost-benefit analyses produced by promoters of major infrastructure projects. Independent studies should be carried out, and, again, such studies should be strong on empirically based risk assessment. Until now it has been difficult or impossible to carry out such assessments, because empirically grounded and statistically valid figures of risk did not exist. With the study documented above this has changed and empirical risk assessment and management has begun (Flyvbjerg, 2006). In addition to sound data, institutional checks and balances that would enforce accountability in actors towards risk are also necessary, as we will see below.

In economics, large inaccuracies in forecasting have recently led to discussions of the necessity of "firing the forecaster" (Akerlof and Shiller 2009, 146). Below, I will argue that this may be letting forecasters off too easily. Some forecasts are so grossly misrepresented that we need to consider not only firing the forecasters but suing them, too, perhaps even having a few serve time.

### IV. EXPLAINING COST OVERRUNS AND BENEFIT SHORTFALLS

Three main types of explanation exist that claim to account for cost overruns and benefit shortfalls in major infrastructure projects: technical, psychological, and political-economic.

Technical explanations account for cost overruns and benefit shortfalls in terms of imperfect forecasting techniques, inadequate data, honest mistakes, inherent problems in predicting the future, lack of experience on the part of forecasters, etc. This is the most common type of explanation of inaccuracy in forecasts (Ascher, 1978; Flyvbjerg, Holm, and Buhl, 2002, 2005; Morris and Hough, 1987; Wachs, 1990). Technical error may be reduced or eliminated by developing better forecasting models, better data, and more experienced forecasters, according to this explanation.

Psychological explanations account for cost overruns and benefit shortfalls in terms of what psychologists call the planning fallacy and optimism bias. Such explanations have been developed by Kahneman and Tversky (1979), Kahneman and Lovallo (1993), and Lovallo and Kahneman (2003). In the grip of the planning fallacy, managers make decisions based on delusional optimism rather than on a rational weighting of gains, losses, and probabilities. They overestimate benefits and underestimate costs. They involuntarily spin scenarios of success and overlook the potential for mistakes and miscalculations. As a result, managers pursue initiatives that are unlikely to come in on budget or on time, or to deliver the expected returns. Overoptimism can be traced to cognitive biases, that is, errors in the way the mind processes information. These biases are thought to be ubiquitous, but their effects can be tempered by simple reality checks, thus reducing the odds that people and organizations will rush blindly into unprofitable investments of money and time.

Political-economic explanations see project planners and promoters as deliberately and strategically overestimating benefits and underestimating costs when forecasting the outcomes of projects. They do this in order to increase the likelihood that it is their projects, and not the competition's, that gain approval and funding. Political-economic explanations have been set forth by Flyvbjerg, Holm, and Buhl (2002, 2005) and Wachs (1989,1990). According to such explanations planners and promoters purposely spin scenarios of success and gloss over the potential for failure. Again, this results in the pursuit of ventures that are unlikely to come in on budget or on time, or to deliver the promised benefits. Strategic misrepresentation can be traced to agency problems and political and organizational pressures, for instance competition for scarce funds or jockeying for position, and it is rational in this sense.[6] If we now define a lie in the conventional fashion as making a statement intended to deceive others (Bok, 1979: 14; Cliffe et al., 2000: 3), we see that deliberate misrepresentation of costs and benefits is lying, and we arrive at one of the most basic explanations of lying that exists: Lying pays off, or at least political and economic agents believe it does. Where there is political pressure there is misrepresentation and lying, according to this explanation, but misrepresentation and lying can be moderated by measures of accountability.

How well does each of the three explanations of forecasting inaccuracy--technical, psychological, and political-economic--account for the data on cost overruns and benefit shortfalls presented earlier? This is the question to be answered below.

Technical explanations have, as mentioned, gained widespread credence among forecasters and project managers (Ascher, 1978; Flyvbjerg, Holm, and Buhl, 2002, 2005). It turns out, however, that such credence could mainly be upheld because until now samples have been too small to allow tests by statistical methods. The data presented above, which come from the first large-sample study in the field, lead us to reject technical explanations of forecasting inaccuracy. Such explanations do not fit the data well. First, if misleading forecasts were truly caused by technical inadequacies, simple mistakes, and inherent problems with predicting the future, we would expect a less biased distribution of errors in forecasts around zero. In fact, we have found with high statistical significance that for four out of five distributions of forecasting errors, the distributions have a mean statistically different from zero. Only the data for inaccuracy in road traffic forecasts have a statistical distribution that seem to fit with explanations in terms of technical forecasting error. Second, if imperfect techniques, inadequate data, and lack of experience were main explanations of inaccuracies, we would expect an improvement in accuracy over time, since in a professional setting errors and their sources would be recognized and addressed through the refinement of data collection, forecasting methods, etc. Substantial resources have in fact been spent over several decades on improving data and methods. Still our data show that this has had no effect on the accuracy of forecasts. Technical factors, therefore, do not appear to explain the data. It is not so-called forecasting "errors" or their causes that need explaining. It is the fact that in a large majority of cases, costs are underestimated and benefits overestimated. We may agree with proponents of technical explanations that it is, for example, impossible to predict for the individual project exactly *which* geological, environmental, or safety problems will appear and make costs soar. But we maintain that it is possible to predict the risk, based on experience from other projects, *that* some such problems will haunt a project and how this will affect costs. We also maintain that such risk can and should be accounted for in forecasts of costs, but typically is not. For technical explanations to be valid, they would have to explain why forecasts are so consistent in ignoring cost and benefit risks over time, location, and project type.

Psychological explanations better fit the data. The existence of optimism bias in managers and promoters would result in actual costs being higher and actual benefits lower than those forecasted. Consequently, the existence of optimism bias would be able to account, in whole or in part, for the peculiar bias found in most of our data. Interestingly, however, when you ask forecasters about causes for forecasting inaccuracies in actual forecasts, they do not mention optimism bias as a main cause of inaccuracy (Flyvbjerg, Holm, and Buhl, 2005: 138-140). This could of course be because optimism bias is unconscious and thus not reflected by forecasters. After all, there is a large body of experimental evidence for the existence of optimism bias (Buehler et al., 1994; Buehler, Griffin, and MacDonald, 1997; Newby-Clark et al. 2000). However, the experimental data are mainly from simple, non-professional settings. This is a problem for psychological explanations, because it remains an open question whether they are general and apply beyond such simple settings. Optimism bias would be an important and credible explanation of underestimated costs and overestimated benefits in infrastructure forecasting if estimates were produced by inexperienced forecasters, i.e., persons who were estimating costs and benefits for the first or second time and who were thus unknowing about the realities of major infrastructure development and were not drawing on the knowledge and skills of more experienced colleagues. Such situations may exist and may explain individual cases of inaccuracy. But given the fact that in modern society it is a defining characteristic of professional expertise that it is constantly tested--through scientific analysis, critical assessment, and peer review--in order to root out bias and error, it seems unlikely that a whole profession of forecasting experts would continue to make the same mistakes decade after decade instead of learning from their actions. Learning would result in the reduction, if not elimination, of optimism bias, which would then result in estimates becoming more accurate over time. But our data clearly shows that this has not happened. The profession of forecasters would indeed have to be an optimistic--and non-professional--group to keep their optimism bias throughout the 70-year period our study covers for costs, and the 30-year period covered for patronage, and not learn that they were deceiving themselves and others by underestimating costs and overestimating benefits. This would account for the data, but is not a credible explanation. Therefore, on the basis of our data, we are led to reject optimism bias as a primary and single cause of cost underestimation and benefit overestimation.

Finally, political-economic explanations and strategic misrepresentation account well for the systematic underestimation of costs and overestimation of benefits found in the data. A strategic estimate of costs would be low, resulting in cost overrun, whereas a strategic estimate of benefits would be high, resulting in benefit shortfalls. A key question for explanations in terms of strategic misrepresentation is whether estimates of costs and benefits are intentionally biased to serve the interests of promoters in getting projects started. This question raises the difficult issue of lying. Questions of lying are notoriously hard to answer, because per definition a lie consists in making a statement intended to deceive others, and in order to establish whether lying has taken place, one must therefore know the intentions of actors. For legal, economic, moral, and other reasons, if promoters and managers have intentionally cooked estimates of costs and benefits to get a project started, they are unlikely to formally tell researchers or others that this is the case. Despite such problems, two studies exist that succeeded in getting forecasters and managers to talk about strategic misrepresentation, one from the UK (Flyvbjerg and Cowi, 2004) and one from the US (Wachs, 1990).

**V. SURVIVAL OF THE UN-FITTEST**
Flyvbjerg and Cowi (2004) interviewed public officials, planners, and consultants who had been involved in the development of large UK transportation infrastructure projects. A planner with a local transportation authority is typical of how respondents explained the basic mechanism of cost underestimation:

> "You will often as a planner know the real costs. You know that the budget is too low but it is difficult to pass such a message to the counsellors [politicians] and the private actors. They know that high costs reduce the chances of national funding."

Experienced professionals like the interviewee know that outturn costs will be higher than estimated costs, but because of political pressure to secure funding for projects they hold back this knowledge, which is seen as detrimental to the objective of obtaining funding.

Similarly, an interviewee explained the basic mechanism of benefit overestimation:

> "The system encourages people to focus on the benefits--because until now there has not been much focus on the quality of risk analysis and the robustness [of projects]. It is therefore important for project promoters to demonstrate all the benefits, also because the project promoters know that their project is up against other projects and competing for scarce resources."

Competition between projects and authorities creates political and organizational pressures that in turn create an incentive structure that makes it rational for project promoters to emphasize benefits and deemphasize costs and risks. A project that looks highly beneficial on paper is more likely to get funded than one that does not.

Specialized private consultancy companies are typically engaged to help develop project proposals. In general, the interviewees found that consultants showed high professional standard and integrity. But interviewees also found that consultants appeared to focus on justifying projects rather than critically scrutinizing them. A project manager explained:

> "Most decent consultants will write off obviously bad projects but there is a grey zone and I think many consultants in reality have an incentive to try to prolong the life of projects which means to get them through the business case. It is in line with their need to make a profit."

The consultants interviewed confirmed that appraisals often focused more on benefits than on costs. But they said this was at the request of clients and that for specific projects discussed "there was an incredible rush to see projects realized."

One typical interviewee saw project approval as "passing the test" and precisely summed up the rules of the game like this:

> "It's all about passing the test [of project approval]. You are in, when you are in. It means that there is so much focus on showing the project at its best at this stage."

In sum, the UK study shows that strong interests and strong incentives exist at the project approval stage to present projects as favorably as possible, that is, with benefits emphasized and costs and risks deemphasized. Local authorities, local developers and land owners, local labor unions, local politicians, local officials, local MPs, and consultants all stand to benefit from a project that looks favorable on paper and they have little incentive to actively avoid bias in estimates of benefits, costs, and risks. National bodies, like certain parts of the Department for Transport and the Ministry of Finance who fund and oversee projects, may have an interest in more realistic appraisals, but so far they have had little success in achieving such realism, although the situation may be changing with the initiatives to curb bias set out in HM Treasury (2003) and UK Department for Transport (2006).

Wachs (1986, 1990) found similar results for transit planning in the US. Taken together, the UK and US studies both account well for existing data on cost underestimation and benefit overestimation. Both studies falsify the notion that in situations with high political and organizational pressure the underestimation of costs and overestimation of benefits is caused by non-intentional technical error or optimism bias. Both studies support the view that in such situations promoters and forecasters intentionally use the following formula in order to secure approval and funding for their projects:

     Underestimated costs
  + Overestimated benefits
  = Funding

Using this formula, and thus "showing the project at its best" as one interviewee said above, results in an inverted Darwinism, i.e., the "*survival of the un-fittest.*" It is not the best projects that get implemented, but the projects that look best on paper. And the projects that look best on paper are the projects with the largest cost underestimates and benefit overestimates, other things being equal. But the larger the cost underestimate on paper, the greater the cost overrun in practice. And the larger the overestimate of benefits, the greater the benefit shortfall. Therefore the projects that have been made to look best on paper in this manner become the worst, or "un-fittest," projects in reality, in the sense that they are the very projects that will encounter most problems during construction and operations in terms of the largest cost overruns, benefit shortfalls, and risks of non-viability. They have been designed like that, as disasters waiting to happen.

**VI. THE OUTSIDE VIEW**
When contemplating what project managers can do to help reform come about, we need to distinguish between two fundamentally different situations: (1) project managers consider it important to get estimates of costs, benefits, and risks right, and (2) project managers do not consider it important to get estimates right, because optimistic estimates are seen as a necessary means to getting projects started. The first situation is the easier one to deal with and here better methodology will go a long way in improving project management. The second situation is more difficult, and more common for political projects as we saw above. Here better governance with changed incentives are essential in order to reward honesty and punish deception, where today's incentives often do the exact opposite.

Thus two main measures of reform are (1) better methods for estimating costs, benefits, and risks, and (2) improved governance. Better methods are covered in this section, improved governance in the next.

If project managers genuinely consider it important to get estimates of costs, benefits, and risks right, it is recommended they use a promising new method called "reference class forecasting" to reduce inaccuracy and bias. This method was originally developed to compensate for the type of cognitive bias in human forecasting that Princeton psychologist Daniel Kahneman found in his Nobel prize-winning work on bias in economic forecasting (Kahneman, 1994; Kahneman and Tversky, 1979). Reference class forecasting has proven more accurate than conventional forecasting. It was used in project management in practice for the first time in 2004 (Flyvbjerg and Cowi, 2004), in 2005 the method was officially endorsed by the American Planning Association (2005), and since then it has been used by governments and private companies in the UK, the Netherlands, Denmark, Switzerland, Australia, and South Africa, among others.

For reasons of space, here I present only an outline of the method, based mainly on Lovallo and Kahneman (2003) and Flyvbjerg (2006). Reference class forecasting consists in taking a so-called "outside view" on the particular project being forecasted. The outside view is established on the basis of information from a class of similar projects. The outside view does not try to forecast the specific uncertain events that will affect the particular project, but instead places the project in a statistical distribution of outcomes from this class of reference projects. Reference class forecasting requires the following three steps for the individual project:

(1) Identification of a relevant reference class of past projects. The class must be broad enough to be statistically meaningful but narrow enough to be truly comparable with the specific project.
(2) Establishing a probability distribution for the selected reference class. This requires access to credible, empirical data for a sufficient number of projects within the reference class to make statistically meaningful conclusions.

(3) Compare the specific project with the reference class distribution, in order to establish the most likely outcome for the specific project.

Daniel Kahneman relates the following story about curriculum planning to illustrate reference class forecasting in practice (Lovallo and Kahneman 2003: 61). Some years ago, Kahneman was involved in a project to develop a curriculum for a new subject area for high schools in Israel. The project was carried out by a team of academics and teachers. In time, the team began to discuss how long the project would take to complete. Everyone on the team was asked to write on a slip of paper the number of months needed to finish and report the project. The estimates ranged from 18 to 30 months. One of the team members--a distinguished expert in curriculum development--was then posed a challenge by another team member to recall as many projects similar to theirs as possible and to think of these projects as they were in a stage comparable to their project. "How long did it take them at that point to reach completion?", the expert was asked. After a while he answered, with some discomfort, that not all the comparable teams he could think of ever did complete their task. About 40 percent of them eventually gave up. Of those remaining, the expert could not think of any that completed their task in less than seven years, nor of any that took more than ten. The expert was then asked if he had reason to believe that the present team was more skilled in curriculum development than the earlier ones had been. The expert said no, he did not see any relevant factor that distinguished this team favorably from the teams he had been thinking about. His impression was that the present team was slightly below average in terms of resources and potential. The wise decision at this point would probably have been for the team to break up, according to Kahneman. Instead, the members ignored the pessimistic information and proceeded with the project. They finally completed the project eight years later, and their efforts went largely wasted--the resulting curriculum was rarely used.

In this example, the curriculum expert made two forecasts for the same problem and arrived at very different answers. The first forecast was the inside view; the second was the outside view, or the reference class forecast. The inside view is the one that the expert and the other team members adopted. They made forecasts by focusing tightly on the case at hand, considering its objective, the resources they brought to it, and the obstacles to its completion. They constructed in their minds scenarios of their coming progress and extrapolated current trends into the future. The resulting forecasts, even the most conservative ones, were overly optimistic. The outside view is the one provoked by the question to the curriculum expert. It completely ignored the details of the project at hand, and it involved no attempt at forecasting the events that would influence the project's future course. Instead, it examined the experiences of a class of similar projects, laid out a rough distribution of outcomes for this reference class, and then positioned the current project in that distribution. The resulting forecast, as it turned out, was much more accurate.

Figure 1 shows what reference class forecasting does in statisticians' language. First, reference class forecasting regresses the best guess of the conventional forecast--here the project promoters' forecast, indicated by the red curve--toward the average of the reference class. The distribution of outcomes in the reference class is indicated by the blue curve. Second, reference class forecasting expands the estimate of interval in the conventional forecast to the interval of the reference class.

[Figure 1 app. here]

With an example from infrastructure provision, planners in a city preparing to build a new subway would, first, establish a reference class of comparable projects. This could be the relevant rail projects from the sample described in Table 1. Through statistical and other analyses the planners would establish which projects to include in the reference class, i.e., projects that were indeed comparable to the one being planned and to each other. Second, if the planners were concerned, for example, with getting construction cost estimates right, they would then establish the distribution of outcomes for the reference class regarding the accuracy of construction cost forecasts. Figure 2 shows what this distribution looks like for a reference class relevant to building subways in the UK, developed by Flyvbjerg and Cowi (2004: 23) for the UK Department for Transport. Third, the planners would compare their subway project to the reference class distribution. This would make it clear to the planners that unless they have reason to believe they are substantially better forecasters and planners than their colleagues who did the forecasts and planning for projects in the reference class, they are likely to grossly underestimate construction costs. Finally, planners would then use this knowledge to adjust their forecasts for more realism. Figure 3 shows what such adjustments are for the UK situation. More specifically, Figure 3 shows that for a forecast of construction costs for a rail project, which has been planned in the manner that such projects are usually planned, i.e., like the projects in the reference class, this forecast would have to be adjusted upwards by 40 percent, if investors were willing to accept a risk of cost overrun of 50 percent. If investors were willing to accept a risk of overrun of only 10 percent, the uplift would have to be 68 percent. For a rail project initially estimated at, say £4 billion, the uplifts for the 50 and 10 percent levels of risk of cost overrun would be £1.6 billion and £2.7 billion, respectively.

[Figures 2 and 3 app. here]

The first instance of reference class forecasting in practice was done in 2004 for capital costs of the proposed Edinburgh Tram Line 2 (Flyvbjerg, 2006). An initial cost estimate of £320 million made by planners was adjusted for optimism bias and acceptable risk, using the probability distribution in Figure 3. This resulted in a new cost estimate of £400 million, including contingencies to insure against cost overruns at the 80 percent level, i.e., with a 20 percent risk of overrun. If the Scottish Parliament, who were underwriting the investment, were willing to accept a risk of overrun of 50 percent, then the cost estimate including contingencies could be lowered to £357 million. Insurance is expensive, here as elsewhere, and the marginal cost of insurance against cost overruns increases as the level of acceptable risk decreases, as seen in Figure 3. Since the Edinburgh Tram, many other infrastructure projects have been subjected to reference class forecasting, Crossrail in London being the most expensive at a cost of £16 billion ($23 billion).

The contrast between inside and outside views has been confirmed by systematic research (Gilovich, Griffin, and Kahneman, 2002). The research shows that when people are asked simple questions requiring them to take an outside view, their forecasts become significantly more accurate. However, most individuals and organizations are inclined to adopt the inside view in planning major initiatives. This is the conventional and intuitive approach. The traditional way to think about a complex project is to focus on the project itself and its details, to bring to bear what one knows about it, paying special attention to its unique or unusual features, trying to predict the events that will influence its future. The thought of going out and gathering simple statistics about related cases seldom enters a manager's mind. This is the case in general, according to Lovallo and Kahneman (2003: 61-62). And it is certainly the case for cost and benefit forecasting in large infrastructure projects. Despite the many forecasts my team and I have reviewed, before the Edinburgh Tram forecast, which is based on our research, we had not come across a single genuine reference class forecast of costs and benefits. Neither had Daniel Kahneman, who first conceived the idea of the reference class forecast.

While understandable, managers' preference for the inside view over the outside view is unfortunate. When both forecasting methods are applied with equal skill, the outside view is much more likely to produce a realistic estimate. That is because it bypasses cognitive and political biases such as optimism bias and strategic misrepresentation and cuts directly to outcomes. In the outside view managers and forecasters are not required to make scenarios, imagine events, or gauge their own and others' levels of ability and control, so they cannot get all these things wrong. Surely the outside view, being based on historical precedent, may fail to predict extreme outcomes, that is, those that lie outside all historical precedents. But for most projects, the outside view will produce more accurate results. In contrast, a focus on inside details is the road to inaccuracy.

The comparative advantage of the outside view is most pronounced for non-routine projects, understood as projects that managers in a certain locale have never attempted before--like building an urban rail system in a city for the first time, or launching a completely new product to the market. It is in the planning of such new efforts that the biases toward optimism and strategic misrepresentation are likely to be largest. To be sure, choosing the right reference class of comparative past projects becomes more difficult when managers are forecasting initiatives for which precedents are not easily found, for instance the introduction of new and unfamiliar technologies. However, many major infrastructure projects are both non-routine locally and use well-known technologies. Such projects are, therefore, particularly likely to benefit from the outside view and reference class forecasting.

**VII. GETTING GOVERNANCE RIGHT**
In the present section we consider the situation where project managers and other influential actors do not find it important to get estimates of costs, benefits, and risks right and where managers, therefore, do not help to clarify and mitigate risks but, instead, generate and exacerbate them. Here project managers are part of the problem, not the solution.

This situation may need some explication, because it possibly sounds to many like an unlikely state of affairs. After all, it may be agreed that project managers and other professionals involved in major infrastructure provision ought to be interested in being accurate and unbiased in their work. It is even stated in the Project Management Institute's Code of Ethics and Professional Conduct (2006: 4, 5) that project managers should "provide accurate information in a timely manner" and they must "not engage in or condone behavior that is designed to deceive others." Economists, engineers, planners, and others involved in major infrastructure provision have similar codes of conduct. But there is a dark side to their work, which is remarkably underexplored in the literature (Flyvbjerg, 1996).

On the dark side, project managers and planners "lie with numbers," as Wachs (1989) has aptly put it. They are busy not with getting forecasts and business cases right and following the PMI Code of Ethics but with getting projects funded and built. And accurate forecasts are often not an effective means for achieving this objective. Indeed, accurate forecasts may be counterproductive, whereas biased forecasts may be effective in competing for funds and securing the go-ahead for a project. "The most effective planner," says Wachs (1989: 477), "is sometimes the one who can cloak advocacy in the guise of scientific or technical rationality." Such advocacy would stand in direct opposition to PMI's ruling that project managers should "make decisions and take actions based on the best interests of society" (Project Management Institute, 2006: 2).

Nevertheless, seemingly rational forecasts that underestimate costs and overestimate benefits have long been an established formula for project approval as we saw above. Forecasting is here mainly another kind of rent-seeking behavior, resulting in a make-believe world of misrepresentation which makes it extremely difficult to decide which projects deserve undertaking and which do not. The consequence is, as even one of the industry's own organs, the Oxford-based Major Projects Association, acknowledges, that too many projects proceed that should not. One might add that many projects don't proceed that probably should, had they not lost out to projects with "better" misrepresentation (Flyvbjerg, Holm, and Buhl, 2002).

In this situation, the question is not so much what project managers can do to reduce inaccuracy and risk in forecasting, but what others can do to impose on project managers the checks and balances that would give managers the incentive to stop producing biased forecasts and begin to work according to their Code of Ethics. The challenge is to change the power relations that govern forecasting and project development. Better forecasting techniques and appeals to ethics won't do here; organizational change with a focus on transparency and accountability is necessary.

As argued in Flyvbjerg, Bruzelius, and Rothengatter (2003), two basic types of accountability define liberal democracies: (1) public sector accountability through transparency and public control, and (2) private sector accountability via competition and the market mechanism. Both types of accountability may be effective tools to curb misrepresentation in project management and to promote a culture which acknowledges and deals effectively with risk, especially where large amounts of taxpayers' money are at stake and for projects with significant social and environmental impacts, as is common with major infrastructure projects.

**Transparency and Public Control**
In order to achieve accountability through transparency and public control, the following would be required as practices embedded in the relevant institutions (the full argument for the measures may be found in Flyvbjerg, Bruzelius, and Rothengatter, 2003, chapters 9-11):

- National-level government should not offer discretionary grants to local agencies for the sole purpose of building a specific type of project (a.k.a. "categorical grants"). Such grants create perverse incentives. Instead, national government should simply offer "block grants" to local governments, and let local political officials spend the funds however they choose to, but make sure that every dollar they spend on one type of project reduces their ability to fund another.

- Cost-benefit analysis and other types of ex ante appraisal should be shifted from promoters to more neutral ground, for instance with the Treasury, in order to reduce risks of agency-problems.
- Forecasts and business cases should be made subject to independent peer review. Where projects involve large amounts of government funds, such review may be carried out by national or state auditing offices, like the General Accounting Office in the US or the National Audit Office in the UK, who have the independence and expertise to produce such reviews.
- Forecasts should be benchmarked against comparable forecasts, for instance using reference class forecasting as described in the previous section.
- For publicly funded projects, forecasts, peer reviews, and benchmarkings should be made available for public scrutiny, including by the media, as they are produced, including all relevant documentation.
- Public hearings, citizen juries, and the like should be organized to allow stakeholders and civil society to voice criticism and support of forecasts. Knowledge generated in this way should be integrated in project management and decision making.
- Scientific and professional conferences should be organized where forecasters would present and defend their forecasts in the face of colleagues' scrutiny and criticism.
- Projects with inflated benefit-cost ratios should be reconsidered and stopped if recalculated costs and benefits do not warrant implementation. Projects with realistic estimates of benefits and costs should be rewarded.
- Professional and occasionally even criminal penalties should be enforced for managers and forecasters who consistently and foreseeably produce deceptive forecasts (Garett and Wachs, 1996).

When I first began suggesting, in lectures for project managers, promoters, and forecasters, that deception and criminal penalties may be concepts relevant to our professions, I would get headshakes, sighs, and the occasional boo. Enron and Iraq changed that, almost overnight. Today people listen and the literature has become replete with books and articles that hammer out the links between lying, forecasting, and management. For instance, a recent book about risk, the planning fallacy, and strategic misrepresentation bluntly states: "Anyone who causes harm by forecasting should be treated as either a fool or a liar. Some forecasters cause more damage to society than criminals" (Taleb, 2007: 163).

Lawmaking has followed suit, most prominently with the 2002 Sarbanes-Oxley Act, which stipulates up to 20 years in prison for a knowingly false forecast intended to impede, obstruct, or influence the proper administration of affairs. There is little doubt that penalties like this influence behavior. The point is that malpractice in project management should be taken as seriously as it is in other professions, e.g., medicine and law. Failing to do this amounts to not taking the profession of project management seriously.

**Competition and Market Control**
In order to achieve accountability via competition and market control, the following would be required, again as practices that are both embedded in and enforced by the relevant institutions:

- The decision to go ahead with a major infrastructure project should, where at all possible, be made contingent on the willingness of private financiers to participate without a sovereign guarantee for at least one third of the total capital needs.[7] This should be required whether projects pass the market test or not, that is, whether projects are subsidized or not or provided for social justice reasons or not. Private lenders, shareholders, and stock market analysts would produce their own forecasts or conduct due diligence for existing ones. If they were wrong about the forecasts, they and their organizations would be hurt. The result would be added pressure to produce realistic forecasts and reduced risk to the taxpayer.
- Forecasters and their organizations must share financial responsibility for covering cost overruns and benefit shortfalls resulting from misrepresentation and bias in forecasting.
- The participation of risk capital would not mean that government reduces control of major infrastructure projects. On the contrary, it means that government can more effectively play the role it should be playing, namely as the ordinary citizen's guarantor for ensuring concerns about safety, environment, risk, and a proper use of public funds.

Whether infrastructure projects are public, private, or public-private, they should be vested in one and only one project organization with a strong governance framework and strong contract-writing skills. The project organization may be a company or not, public or private, or a mixture. What is important is that this organization has the capacity to (a) set up and negotiate contracts that will effectively safeguard its interests, including in equity risk allocation, and (b) enforce accountability vis-à-vis contractors, operators, etc. In turn, the directors of the organization must be held accountable for any cost overruns, benefit shortfalls, faulty designs, unmitigated risks, etc. that may occur during project planning, implementation, and operations.

Experience with contract writing is a much-neglected topic, but is particularly important in developing and managing major infrastructure projects. This is because a fundamental asymmetry in experience with and resources allocated to contract writing often applies in the client-contractor relationship for such projects. Clients who decide to do major infrastructure--for instance a city council deciding to build a new subway or toll road--do so relatively rarely, often only once, or never, in the lifetime of the individual city manager and council member. Learning is therefore impaired for clients, and if you don't know what your interests are, it's difficult to safeguard them. Contractors, on the other hand, who bid for and build such projects, do so all the time. Contractors therefore typically know much more than clients about the ins and outs of projects and contracts, including the many risks and pitfalls that apply, plus which lawyers, bankers and consultants to hire to safeguard their interests most effectively. This asymmetry has brought many a client to grief. A possible way to bring more symmetry into the client-contractor relationship would be for government to establish a central contract-writing unit at the state or national level, which would be in charge of negotiating, on behalf of local and other branches of government, the types of major contracts they do too infrequently to gain real experience. This would concentrate a larger number of contracts in one place, allowing experience--and the negotiating power that comes with it--to accumulate.

Fortunately, better governance along the lines described above has recently become stronger around the world. The Enron scandal and its successors have triggered new legislation and a war on corporate deception that is spilling over into government with the same objective: to curb financial waste and promote good governance. Although progress is slow, good governance is gaining a foothold also in major infrastructure project management.

For example, in 2003 the Treasury of the United Kingdom required, for the first time, that all ministries develop and implement procedures for major projects that will curb what the Treasury calls--with true British civility--"optimism bias." Funding will be unavailable for projects that do not take into account this bias, and methods have been developed for how to do this (HM Treasury, 2003; Flyvbjerg and Cowi, 2004; UK Department for Transport, 2006). In the Netherlands in 2004, the Parliamentary Committee on Infrastructure Projects for the first time conducted extensive public hearings to identify measures that will limit the misinformation about large infrastructure projects given to the Parliament, public, and media (Tijdelijke Commissie Infrastructuurprojecten, 2004). In Boston, the government has sued to recoup funds from contractor overcharges for the Big Dig related to cost overruns. More countries and cities are likely to follow the lead of the UK, the Netherlands, and Boston in coming years; Switzerland and Denmark are already doing so (Swiss Association of Road and Transportation Experts, 2006; Danish Ministry for Transport and Energy, 2006, 2008).

Moreover, with private finance in major infrastructure projects on the rise over the past 15-20 years, capital funds and banks are increasingly gaining a say in the project development and management process. Private capital is no panacea for the ills in major infrastructure project management, to be sure (Hodge and Greve, 2009). But private investors place their own funds at risk, as opposed to governments who place the taxpayer's money at risk. Capital funds and banks can therefore be observed to not automatically accept at face value the forecasts of project managers and promoters. Banks typically bring in their own advisers to do independent forecasts, due diligence, and risk assessments, which is an important step in the right direction. The false assumption that one forecast or one business case (which is also a forecast) may contain the truth about a project is problematized. Instead project managers and promoters are getting used to the healthy fact that different stakeholders have different forecasts and that forecasts are not only products of objective science and engineering but of negotiation. Why is this more healthy? Because it is more truthful about our ability to predict the future and about the risks involved.

If the institutions with responsibility for developing and building major infrastructures would continue to effectively implement, embed, and enforce such measures of accountability, then the misrepresentation in cost, benefit, and risk estimates, which is widespread today, may be mitigated. If this is not done, misrepresentation is likely to continue, and the allocation of funds for major infrastructure is likely to continue to be wasteful, unethical, and sometimes even unlawful.

**VIII. CURRENT TRENDS IN INFRASTRUCTURE INVESTMENT**
In the introduction to this article, I mentioned that current spending on infrastructure constitutes the biggest investment boom in history, measured as share of world GDP. We also saw above that even in the best of times large infrastructure investments have a dismal performance record in terms of cost overruns, delays, and benefit shortfalls. Nine out of ten projects experience cost overrun, and overrun has not diminished for the 70 years for which we have data, to mention but two grim statistics.

Throwing hundreds of billions of extra stimulus dollars at an underperforming business that is already at bubble-like investment levels, is therefore highly risky at best. Nevertheless, this is what China, the US, and many other countries decided to do with their stimulus packages in 2008 and 2009. Risks include rampant pork-barrel, fast-tracking, bid rigging, local governments pulling their funds out of on-going projects in anticipation of national funding that may not come or comes late, and projects left unfinished because of cost overruns on stimulus projects that local government cannot finance. The consequences could be dire to the economy, and to public trust in the institutions and people who administer infrastructure spending. Perhaps this is why Macquarie Bank--probably the largest and most experienced infrastructure investor in the world--began reducing its infrastructure portfolio in 2009, moving into energy instead.

Add to this that global infrastructure investments are shifting geographically from developed to emerging economies, where risks of ineffective project delivery are even higher. Over half of infrastructure investments are now taking place in emerging economies. In the past five years, China has spent more on infrastructure in real terms than in the whole of the 20th century. In the past four years, China has built as many kilometers of high-speed passenger rail lines as Europe has in two decades. Morgan Stanley predicts that emerging economies will spend $22 trillion in today's prices on infrastructure over the next ten years. Political risks, risks of corruption, and logistical risks are higher in emerging economies. As a consequence, risks of cost overruns and benefit shortfalls are also higher. A recent study documented that the average cost overrun on rail projects--measured in real terms from the decision date to completion-- was 64.6 percent in emerging economies against 40.8 percent in North America and 34.2 percent in Europe (Flyvbjerg, Holm, and Buhl, 2002: 285). Geography matters to project performance. The striking shift in infrastructure investments to emerging economies is therefore placing increased pressure on project delivery.

Finally, information and communication technology (ICT) has fast become a large and rapidly increasing part of more major infrastructure projects. The consequences are often devastating, because large ICT projects appear to perform even worse than other major projects. Recently, a group of ICT experts contacted me to ask how they might learn to manage cost overruns, delays, etc. in ICT projects "as well as in transportation infrastructure projects." I laughed out loud because I mistakenly thought they were making a joke. As mentioned above, the evidence is clear that most major transportation infrastructure projects perform poorly. But here is another project type, ICT, that apparently performs so much worse that it can use transportation as a benchmark to be strived for. We did a pilot study, and sure enough, the ICT experts were right: If a major project is not already messed up, injecting a good dose of ICT will do the job (see Figure 4).

[Figure 4 app. here]

As if it were not difficult enough to develop, say, a major new airport, we are now developing airports that depend on major new ICT for their operations, and we pay the price. Hong Kong international airport is a case in point, with initial ICT hiccups so bad that the whole Hong Kong economy suffered. Terminal 5 at London Heathrow is another example. An infrastructure planner recently told me, "We know how to build large, expensive tunnels by now, but we don't know how to build the ICT safety systems that go into the tunnels; ICT busts us every time."

Or take Germany's Toll Collect, a consortium of DaimlerChrysler, Deutsche Telekom, and Cofiroute of France responsible for tolling heavy trucks on German motorways for the Federal government. The new tolling system was designed to be a showcase for public-private partnership in infrastructure management. Tolling was scheduled to start in January 2003. A year later the project was falling apart. The developers had been too optimistic about the software that would run the system. The government was losing toll revenues estimated to total €6.5 billion ($8.6 billion). For lack of funds, all new road projects in Germany and related public works were put on hold, threatening 70,000 construction jobs. Politicians and media were calling for prosecution of Toll Collect for deceiving the government. Finally, the German transport minister cancelled the contract with Toll Collect and gave the company two months to come up with a better plan. By the time tolling at last started, several years delayed, "Toll Collect" had become a popular byword among Germans used to describe everything wrong with the national economy. Projects like this will not get us out of the current crisis. Quite the opposite.

Examples abound and the advice regarding ICT seems obvious: If you are doing ICT as part of a major project (or as a major project in itself) be sure to get ICT that has been developed and debugged elsewhere and that has a proven track record in daily use. If you decide not to follow this advice (which is common) you should be explicit that you are in essence taking on the development risk of a major new product, which translates into large and unpredictable risks of delays, cost overruns, and benefit shortfalls. Your set-up for risk management and contingencies should then duly reflect this (which is uncommon).

And here's the crunch: Not only are clients and managers of major infrastructure projects rushing headlong into risky ICT investments that are really new product development schemes, but a rapidly increasing part of these investments are shifting to emerging economies, which is a source of risk in its own right, as we saw above. Between 2003 and 2008, the share of ICT spending in emerging economies rose by 60 percent, from 15 to 24 percent of global ICT spending, according to the OECD.

In sum, the following formula captures the current trends in major infrastructure investment:

```
   (rapid increase in infrastructure spending)
 + (rapid shift in spending to emerging economies)
 + (rapid increase in spending on ICT)
 = (Risk to the third degree of financial and economic disasters)
```

Whether stimulus infrastructure spending will improve or worsen the global economy will be decided by how well we deal with these main trends. Thus the efficiency of project delivery is both particularly important at present and particularly challenged. The major challenges are (a) to not lower standards as the project pipeline rapidly expands, (b) to navigate the particular risks of doing more projects in emerging economies, and (c) to harness the bull in a china shop that is ICT in major projects.

Three main ingredients will help meet the challenges. First, we need to honestly acknowledge that infrastructure investment is no easy fix but is fraught with problems. For this purpose, President Obama was immensely helpful when at a 2009 White House Fiscal Responsibility Summit he openly identified "the costly overruns, the fraud and abuse, the endless excuses" in public procurement as key problems (The White House, 2009). *The Washington Post* (February 24, 2009) rightly called this "a dramatic new form of discourse." Before Obama it was not *comme-il-faut* to talk about overruns, deception, and abuse in relation to infrastructure spending, although they were of epidemic proportions then as now, and the few who did use such language were ostracized. However, we cannot solve problems we cannot talk about. So talking is the first step.

Second, we must arrive at a better understanding and better management of the long, fat tails of financial and economic risks--the abundance of black swans--that apply to infrastructure investment. Risks have so far been as misunderstood and as mismanaged in infrastructure investment as in the financial markets, with equally devastating outcomes. Methods and data for better risk assessment and management were presented above.

Third, incentives need to be put straight, so that bad performance is punished and good performance rewarded, and not the other way around, which is often the case today. Again, methods for how to do this were described above.

**IX. CONCLUSIONS**
This article documents a much neglected topic in economics, namely the fact that ex ante estimates of ventures' costs and benefits are often very different from actual ex post costs and benefits. The article shows that such differences between estimated and actual outcomes are pronounced for large infrastructure projects, where substantial cost underestimates often combine with equally significant benefit overestimates, rendering cost-benefit analyses of projects not only inaccurate but biased.

The cause of biased cost-benefit analyses is found to be perverse incentives that encourage promoters of infrastructure projects to underestimate costs and overestimate benefits in the business cases for their projects in order to gain approval and funding. But the projects that are artificially made to look best in business cases are the projects that generate the highest cost overruns and benefit shortfalls in reality, resulting in a significant trend for "survival of the un-fittest" for infrastructure projects.

The cure to the problem is enforcing an outside view in the planning of new projects and employing a method called reference class forecasting, based on Daniel Kahneman's Nobel Prize-winning theories of decision making under uncertainty. However, to be effective such new methodology must be combined with better governance structures with incentives that reward accurate estimates of costs and benefits and punish inaccurate ones.

Finally, stimulus spending has recently resulted in extra money and attention for infrastructure investing. This is placing increased pressure on project delivery. Stimulus spending--together with rapidly increasing spending on infrastructure in emerging economies and on information technology in infrastructure--is driving infrastructure investment from the frying pan into the fire.

Table 1: Inaccuracy of transportation project cost estimates by type of project, in constant prices.

| Type of project | No. of cases (N) | Avg. cost overrun % | Standard deviation |
|---|---|---|---|
| Rail | 58 | 44.7 | 38.4 |
| Bridges and tunnels | 33 | 33.8 | 62.4 |
| Road | 167 | 20.4 | 29.9 |

Table 2: Inaccuracy in forecasts of rail passenger and road vehicle traffic.

| Type of project | No. of cases (N) | Avg. inaccuracy % | Standard deviation |
|---|---|---|---|
| Rail | 25 | −51.4 | 28.1 |
| Road | 183 | 9.5 | 44.3 |

Figure 1: What reference class forecasting does, in statisticians' language.

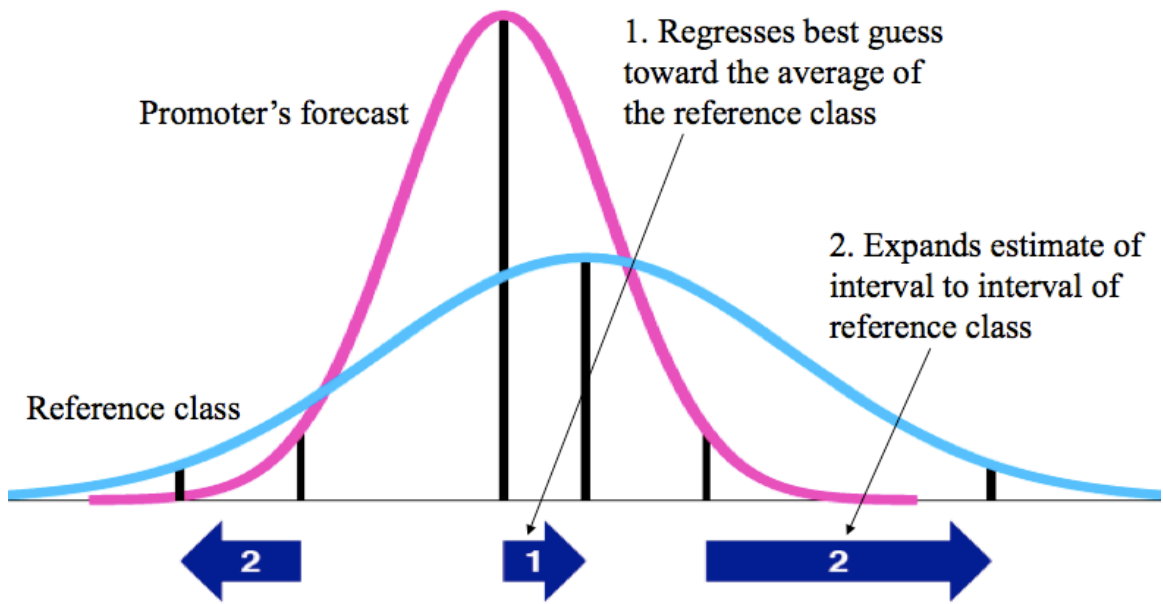

Figure 2: Inaccuracy of construction cost forecasts for rail projects in reference class. Average cost increase is indicated for non-UK and UK projects, separately. Constant prices.

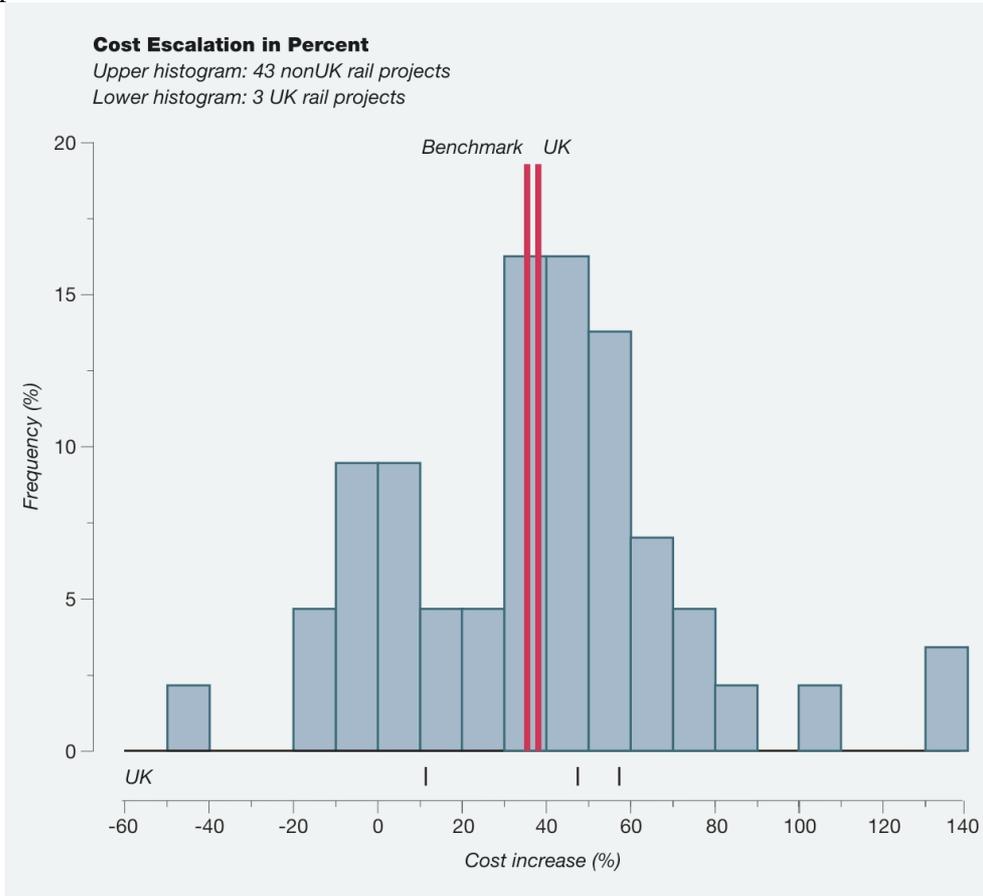

Figure 3: Required adjustments to cost estimates for UK rail projects as function of the maximum acceptable level of risk for cost overrun. Constant prices.

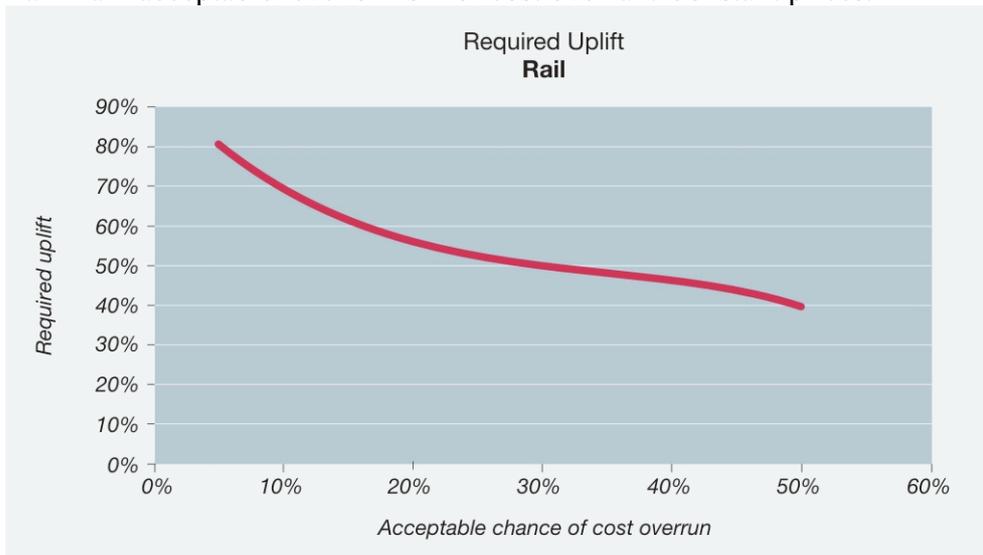

Figure 4: Cost overrun in construction projects and IT projects compared.

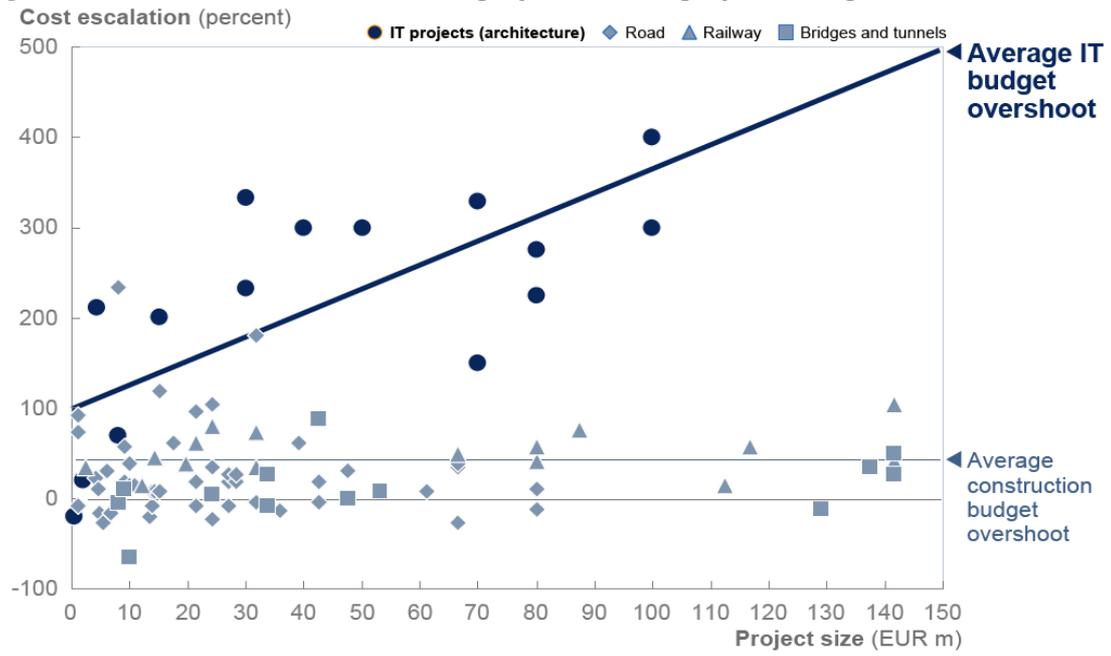

## Notes

[1] By "major projects" I here mean the most expensive projects that are built in the world today, typically at costs per project from around a hundred million to several billion dollars.

[2] The data are from the largest database of its kind. All costs are construction costs measured in constant prices. Cost overrun, also sometimes called "cost increase" or "cost escalation," is measured according to international convention as actual out-turn costs minus estimated costs in percent of estimated costs. Actual costs are defined as real, accounted construction costs determined at the time of project completion. Estimated costs are defined as budgeted, or forecasted, construction costs at the time of decision to build. For reasons explained in Flyvbjerg, Holm, and Buhl (2002) the figures for cost overrun presented here must be considered conservative. Ideally financing costs, operating costs, and maintenance costs would also be included in a study of costs. It is difficult, however, to find valid, reliable, and comparable data on these types of costs across large numbers of projects. For details on methodology, see Flyvbjerg, Holm, and Buhl (2002).

[3] Following international convention, inaccuracy is measured as actual traffic minus estimated traffic in percent of estimated traffic. Rail traffic is measured as number of passengers; road traffic as number of vehicles. The base year for estimated traffic is the year of decision to build. The forecasting year is the first full year of operations. Two statistical outliers are not included here. For details on methodology, see Flyvbjerg (2005b).

[4] For each of twelve urban rail projects, we have data for both cost overrun and traffic shortfall. For these projects average cost overrun is 40.3 percent; average traffic shortfall is 47.8 percent.

[5] Quoted from "Under Water Over Budget," *The Economist*, October 7, 1989, 37–38.

[6] For an interpretation of strategic misrepresentation in terms of agency problems and rent-seeking behavior, see Flyvbjerg, Garbuio**,** and Lovallo (2009).

[7] A sovereign guarantee is a guarantee where government takes on the risk of paying back a loan, even if the loan was obtained in the private lending market. The lower limit of a one-third share of private risk capital for such capital to effectively influence accountability is based on practical experience. See more in Flyvbjerg, Bruzelius, and Rothengatter (2003: 120-123).